# Optical synthetic sampling imaging: concept and an example of microscopy

**JUNZHENG PENG,**[1,2,3] **MANHONG YAO,**[1,3] **ZIXIN CAI,**[1] **XUE QIU,**[1] **ZIBANG ZHANG,**[1,2] **SHIPING LI,**[1,2] **AND JINGANG ZHONG**[1,2,*]

[1]*Department of Optoelectronic Engineering, Jinan University, Guangzhou 510632, China*
[2]*Guangdong Provincial Key Laboratory of Optical Fiber Sensing and Communications, Jinan University, Guangzhou 510632, China*
[3]*These authors contributed equally*
*\* Corresponding author: tzjg@jnu.edu.cn*

**Abstract:** Digital two-dimensional (2D) spatial sampling devices (such as charge-coupled device) have been widely used in various imaging systems, especially in computational imaging systems. However, the undersampling of digital sampling devices is a problem that limits the resolution of the acquired images. In this study, we present a synthetic sampling imaging (SSI) concept to solve the undersampling problem. It combines the structured illumination system and conventional 2D image detection system to simultaneously sample the specimen from the illumination and the detection sides. Then, we synthesize the illumination sampling rate and the detection sampling rate to reconstruct a high sampling rate image. The concept of the proposed SSI is demonstrated by an example of microscopy. Experimental results confirm that the proposed method can double the sampling resolution of the microscope. The synthetic sampling scheme, where the sampling task is shared by the illumination and detection sides, provides insight for resolving the undersampling problem of the digital imaging system.

## 1. Introduction

Digital imaging is essentially a two-step spatial information acquisition process. The first step involves producing an analog image of the object through a lens, while the second step involves sampling the analog image using a two-dimensional (2D) image sensor to produce the final digital image. However, spatial information encounters a certain level of loss in each step. In the first step, the lens system is a low-pass filter whose cutoff frequency is given by its numerical aperture (NA) according to the Abbe's theory [1]. In the second step, the 2D image sensor is also a low-pass filter whose cutoff frequency is given by its sampling rate (namely, the inverse of the pixel pitch) according to the Nyquist-Shannon sampling theory. The overall cutoff frequency of a digital imaging system is determined by the lower cutoff frequency between the utilized lens system and sensor. For visible light imaging, modern semiconductor technologies allow us to manufacture high-performance charge-coupled device (CCD) and complementary metal-oxide semiconductor (CMOS)-based cameras with a microscale pixel pitch. In this case, the cutoff frequency of the imaging lens is lower than that of the 2D imaging sensor, one can use super-resolution imaging techniques to improve the cutoff frequency of the imaging lens [2-5]. However, for non-visible light imaging, manufacturing a 2D sensor with the corresponding resolution remains challenging. Additionally, a smaller pixel pitch yields a lower signal-to-noise ratio. In this case, the cutoff frequency of the 2D imaging sensor is lower than that of the imaging lens. As such, we may encounter the undersampling problem in many imaging systems, especially in non-visible light imaging systems.

Recent developments in single-pixel imaging techniques provide a solution for reconstructing a digital image without using a 2D image sensor [6-14]. Single-pixel imaging techniques use a group of illumination patterns (e.g., Hadamard or Fourier basis patterns) generated by a spatial light modulator (SLM) to encode the 2D spatial information of an object into a one-dimensional (1D) time-varying intensity signals [15-22], and then use a spatial non-resolved detector to collect the time-varying light intensity signals. Clearly, the spatial information of the object is sampled by the illumination patterns. The highest frequency of the illumination patterns determines the resolution of the reconstructed image [7]. To acquire a high-sampling rate image via the single-pixel imaging technique, we must use a high-sampling rate SLM. However, for the non-visible light wavebands, e.g., the terahertz wave, it is difficult to fabricate a high-sampling rate SLM for the structured illumination [23]. Thus, non-visible light imaging remains challenging because the single-pixel imaging also suffers from the undersampling problem.

To solve the undersampling problem, we propose an optical synthetic sampling imaging (SSI) technique. The concept of the optical SSI is inspired by the optical synthetic aperture imaging [24]. To break the resolution limit imposed by the NA of the imaging lens, the synthetic aperture imaging technique synthetizes the NAs of the illumination lens and imaging lens to increase the NA of the imaging system. Similar to the synthetic aperture imaging system, the proposed SSI system is a combination of the structured illumination system and the conventional imaging system. It not only samples specimen from the illumination side, but also samples specimen from the detection side. We consider the SSI can improve the sampling rate of the imaging system because it synthetizes the illumination sampling rate and detection sampling rate. If so, we can resolve the undersampling problem.

We use an example of microscopic imaging called synthetic sampling microscopy to demonstrate the feasibility of the SSI method. With it, we can simultaneously obtain multiple low-sampling rate images using the synthetic sampling microscope. These low-sampling rate images are shifted with respect to each other via sub-pixel displacements. Based on these shifted images, we can synthesize a high-sampling rate image using the pixel super-resolution algorithm [25]. The experimental results show that the proposed SSI method is effective.

## 2. Principle

### 2.1 Undersampling problem

Figure 1(a) shows a conventional light microscope, where a parallel light beam is used to illuminate the specimen, and the spatial information of the specimen is sampled by a 2D sensor. Assume that the spatial cutoff frequency determined by the objective can be represented as $f_{obj}$ and the spatial cutoff frequency determined by the 2D sensor can be represented as $f_{cam}$. If $f_{cam} < f_{obj}$, the imaging resolution is determined by the sampling rate of the 2D sensor, which leads to the undersampling problem. A common way to solve this problem is to obtain multiple sub-pixel shifting images and synthesize these images with the pixel super-resolution algorithm [25]. To acquire multiple sub-pixel shifting images, however, this usually requires mechanical scanning for the specimen, which significantly increases the measurement time.

Figure 1(b) shows a single-pixel microscope. It uses a set of structured illumination patterns to sample the specimen, and then uses a non-spatially resolved detector to collect the light signals. Assume that the spatial cutoff frequency determined by the objective lens is represented as $f_{ill}$, and the spatial cutoff frequency determined by the SLM is represented as $f_{SLM}$. If $f_{SLM} < f_{ill}$, the imaging resolution of the single-pixel microscope is determined by the sampling rate of the SLM, which results in the undersampling problem. Similar to the case of conventional microscope, the undersampling problem of the single-pixel microscope can be alleviated by using micro-scanning methods [26-29]. However, these methods require mechanical scanning for the SLM or require shifting the illumination patterns, significantly increasing the measurement time.

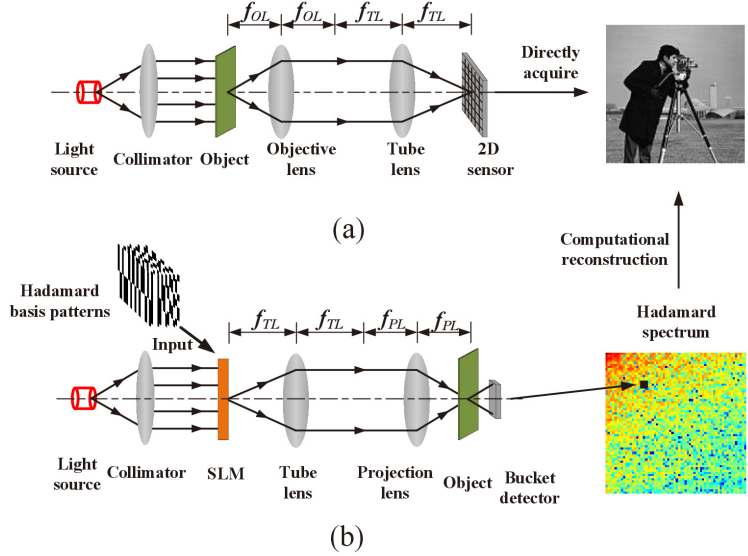

Fig. 1. (a) In conventional light microscopy, an image is acquired by sampling object information from the detection side; (b) single-pixel microscope reconstructs an image by sampling object information from the illumination side. ($f_{TL}$: focal length of the tube lens; $f_{OL}$: focal length of the objective lens; $f_{PL}$: focal length of the projection objective lens).

In short, we can retrieve an image by sampling the spatial information of the specimen from either the detection side or the illumination side. To satisfy the Nyquist-Shannon sampling theorem and avoiding undersampling, the existing methods usually require a high-sampling rate 2D sensor or SLM. This can be readily realized in the visible light imaging system. However, for non-visible light imaging, fabricating a high-sampling rate 2D sensor or SLM is difficult or significantly more expensive. Although the existing pixel super-resolution methods can be used to alleviate the undersampling problem, they require a mechanical scanning for the acquisition of multiple sub-pixel shifting images.

*2.2 Synthetic sampling microscopy*

To overcome the undersampling problem, we propose an SSI technique. The SSI system is a combination of the structured illumination and conventional 2D image detection systems. By simultaneously sampling the spatial information of the specimen from the detection and illumination sides, we can synthesize the illumination and detection samplings. Thus, the sampling rate of the imaging system can be improved.

To realize SSI using the conventional microscope, we employ a 2D array of light-emitting diodes (LEDs) in the illumination side, as shown in Figs. 2(a-c). For simplicity, we assume that the specimen (denoted as black arrow) contains only one layer. Then, we can draw the following conclusions from Figs. 2(a-c):

- If the specimen is placed at the focal plane of the objective lens, all images will be located at the same position whatever it is illuminated with an on-axis LED or off-axis LED, as shown in Fig. 2(a).
- If the specimen is placed slightly away from the focal plane of the objective lens but within the depth of field of the objective lens, as shown in Figs. 2(b-c), the image of the specimen will be laterally shifted with the position of LED. Noted that the resolution of the image will not be reduced when the specimen is placed within the depth of field of the objective lens, compared with that imaging at the focal plane.

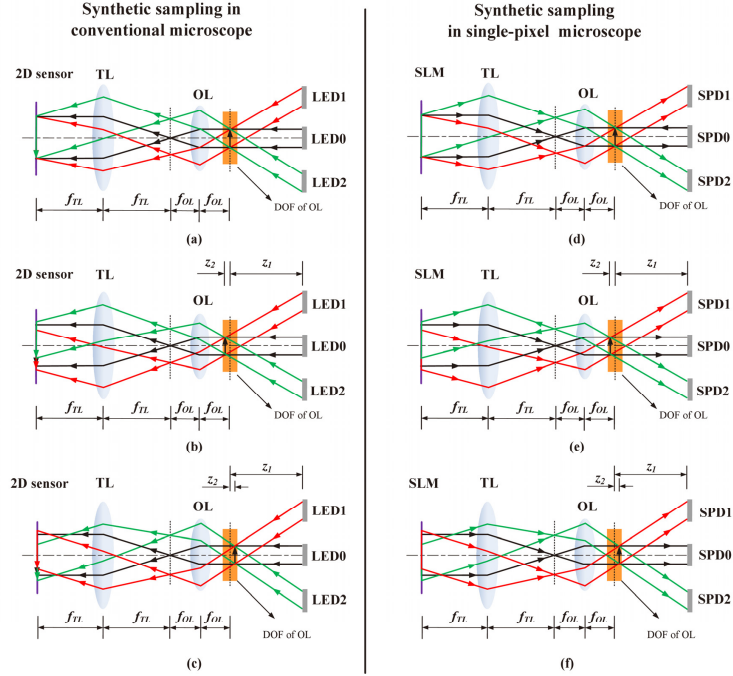

Fig. 2. Principle of synthetic sampling imaging. (a-c) Synthetic sampling using the conventional microscope; (d-f) synthetic sampling using the single-pixel microscope. (TL: tube lens; OL: objective lens; DOF: depth of field; SPD: single-pixel detector; $f_{TL}$: focal length of the tube lens; $f_{OL}$: focal length of the objective lens.)

From the above analysis, it can be concluded if we move the specimen slightly away from the focal plane but within the depth of field of the objective lens, we would obtain multiple images that are shifted with respect to each other, without using mechanical scanning. Based on these shifting images, we can synthesize a high-sampling rate image via the pixel super-resolution algorithm. Thus, the undersampling problem of the 2D sensor can be resolved using the proposed synthetic sampling method.

Alternatively, to realize the synthetic sampling imaging using the single-pixel microscope, we employ an array of single-pixel detectors (as shown in Figs. 2(d-f)) instead of a single-pixel detector to collect the light signals from the detection side. With the single-pixel imaging method, the 2D spatial information of the specimen is encoded into a 1D time-varying light signals. We assume that the illumination source given in Figs. 2(d-f) is divergent, and the specimen has property of weak scattering. Then the 2D spatial information of the specimen will be simultaneously modulated by the structured patterns from different directions, such as the red, green, and black light rays shown in Figs. 2(d-f). As such, multiple single-pixel detectors located at different positions can simultaneously record the 1D light signals, and then generate multiple 2D images simultaneously [22].

More interestingly, when we move the specimen from the focal plane with a small distance but within the depth of field of the objective lens, the multiple images retrieved from the array of single-pixel detectors will also be shifted with respect each other. Because according to the principle of Helmholtz reciprocity [7, 8, 22], the imaging systems given in Figs. 2(d-f) are equivalent to the configurations shown in Figs. 2(a-c). Specifically, the SLM shown in Figs. 2(d-f) is equivalent to the 2D sensor shown in Figs. 2(a-c). Each single-pixel detector shown in Figs. 2(d-f) is equivalent to the LED source shown in Figs. 2(a-c). If we move the specimen from the focal plane with the same distance as that shown in Figs. 2(b-c), according to Helmholtz reciprocity, the multiple images retrieved by the imaging system given in Figs. 2(e-f) will also be shifted with respect to each other. In addition, because all single-pixel detectors can simultaneously work, we can simultaneously retrieve multiple images by using an array of single-pixel detectors. Based on these shifting images, we can synthesize a high-sampling rate image via the pixel super-resolution algorithm. Therefore, the undersampling problem of the SLM can also be resolved with the proposed synthetic sampling method.

### 3. Determination of sub-pixel shifting amount and reconstruction of high-sampling rate image

To synthesize a high-sampling rate image, the low-sampling rate images should be shifted with respect to each other via sub-pixel displacements. To achieve this using the system shown in Figs. 2(a-c), we should carefully set the axial distance (denoted as $z_1$) between the LED and the focal plane of the objective lens. Assume that $z_2$ is the axial distance between the specimen and the focal plane of the objective, and $S_{LED}$ is the lateral distance between LEDs. Using geometric knowledge, the ratio between the shift of the retrieved image (denoted as $S_{image}$) and $S_{LED}$ can be approximated as follows [30]:

$$\frac{S_{image}}{M \cdot S_{LED}} = \frac{z_2}{z_1}, \tag{1}$$

where $M$ is the magnification of microscope. If $z_1 \gg z_2$, $z_2/z_1$ is exceedingly small. By adopting this demagnification feature, we can readily retrieve a sub-pixel shifting image by setting $S_{LED}$ as several millimeters. For example, if $z_1 = 50$ mm, $z_2 = 0.01$ mm, and $M=10$, the ratio between the two shifts is $S_{image}/(M \bullet S_{LED}) = 1/5000$. Assume that the pixel size of the retrieved image is 10 μm. To acquire multiple images that are shifted with respect to each other via subpixel displacement (e.g., 2 μm), we must set $S_{LED}$ as 1 mm. This can be readily realized. In contrast to existing methods, the proposed method avoids mechanical scanning. Because the measurement systems illustrated in Figs 2(a-c) and 2(e-f) are subject to Helmholtz reciprocity, to retrieve multiple sub-pixel shifting images via the measurement system shown in Figs. 2(e-f), we can set the systems shown in Figs. 2(e-f) with Eq. (1).

Before synthesizing the high-sampling rate image, we should determine the sub-pixel shift for every low-sampling rate image. To realize this, we take two different low-sampling rate images retrieved from the SSI system as an example. Suppose that one image is the template image (denoted as $T(x)$), and the other image is the warped image (denoted as $I(x)$). $x = (x, y)^T$ is a column vector containing the pixel coordinates. Additionally, suppose that $p = (p_1, p_2)^T$ is a column vector representing the shift amount between the two images. According to the rigid transformation, we can map the points of template image to the warp image using the following expression,

$$W(x; p) = \begin{bmatrix} x + p_1 \\ y + p_2 \end{bmatrix}, \tag{2}$$

where $W(x; p)$ represents the pixel coordinates of a point in the warped image that corresponds to a point in the template image. To estimate the vector $p$, we define an objective function $O(p)$ by using the difference or the dissimilarity of the intensity values of the correspondences

$$O(p) = \sum_{i=1}^{N} \left[ I(W(x; p)) - T(x) \right]^2. \tag{3}$$

Clearly, the problem of estimating the parameter vector $p$ involves minimizing the sum of squared dissimilarity between the two images. Existing methods attempt to minimize Eq. (3) by using direct search or the gradient-based approach. Because the gradient-based approach has low computational cost, we use it to solve the parameter vector $p$. Assume that $\tilde{p}$ denotes the estimated values of the parameter vector $p$, and $p = \tilde{p} + \Delta p$. The difference or the dissimilarity between the low-sampling rate images is very small. By applying the first-order Taylor expansion with respect to the parameter vector $p$, Eq. (3) can be rewritten as follows

$$O(p) = \sum_{x}\left[I(W(x;\tilde{p}+\Delta p)) - T(x)\right]^2$$
$$= \sum_{x}\left[I(W(x;\tilde{p})) + \nabla I \frac{\partial W}{\partial p}\Delta p - T(x)\right]^2, \quad (4)$$

where $\nabla I = \left(\frac{\partial I}{\partial x}, \frac{\partial I}{\partial y}\right)$ is the gradient of image $I$ evaluated at $W(x,\tilde{p})$, and $\frac{\partial W}{\partial p}$ is the Jacobian of the warp. Clearly, Eq. (4) is a least-squares problem. We can solve $\Delta p$ from Eq. (4) using the steepest-decent algorithm [31]:

$$\Delta p = H^{-1} \sum_{x}\left[\nabla I \frac{\partial W}{\partial p}\right]^T \left[T(x) - I(W(x;\tilde{p}))\right]. \quad (5)$$

Here, $H$ is the Hessian matrix:

$$H = \sum_{x}\left[\nabla I \frac{\partial W}{\partial p}\right]^T \left[\nabla I \frac{\partial W}{\partial p}\right]. \quad (6)$$

According to Eqs. (2-6), we propose a three-step iterative method for estimating the sub-pixel shifting amount for each low-sampling rate image. First, initialize $\tilde{p}^{(0)} = [0,0]^T$. Then use Eqs. (4-6) to calculate $\Delta p$. Second, update $\tilde{p}$ with the following expression

$$\tilde{p}^{(k+1)} = \tilde{p}^{(k)} + \Delta p, \quad (7)$$

where $\tilde{p}^{(k+1)}$ and $\tilde{p}^{(k)}$ are the solutions of the kth and (k+1)th iterations, respectively. Third, step two is repeated until $\Delta p < \varepsilon$, where $\varepsilon$ is a user-defined threshold. In practice, each low-sampling rate image is retrieved under a different illumination direction or from a different view; thus, the contrast and brightness might differ among the images. To consider these effects, the reader can use the enhanced correlation coefficient (ECC) algorithm [32] to estimate the parameter vector $p$. The ECC algorithm has been developed by G. Evangelidis using MATLAB code, and the reader can download it from [33]. To calculate the sub-pixel shifting amounts for all low-sampling rate images, we can arbitrarily choose one low-sampling rate image as the template image and use the aforementioned method to calculate the sub-pixel shifting amounts between the template image and others images.

After obtaining the sub-pixel shifts for all low-sampling rate images, we can use the pixel super-resolution algorithm to synthesize a high-sampling rate image. The pixel super-resolution method usually forms a group of linear equations to represent the mapping relationship between the low-sampling rate images and the high-sampling rate image [34]

$$L = PH + E, \quad (8)$$

where $L$ concatenates column vectors formed by all low-sampling rate image, $H$ includes all pixels of the high-sampling rate image, $P$ is a weighting matrix that maps the high-sampling rate image to the low-sampling rate image, and $E$ is the noise. By using the method given in appendix A, we can obtain the weighting matrix $P$. Finally, the high-sampling rate image can be solved from Eq. (8).

## 4. Experiments

To demonstrate our method, we built a synthetic sampling microscope via single-pixel imaging. As shown in Fig. 3, light coming from an LED source (center wavelength: 633 nm) was directed onto a digital micromirror device (DMD) ($9.5''$; $1{,}920\times1{,}080$ pixels; pixel size: 10.8 μm) by a reflecting mirror. Because single-pixel imaging using Hadamard patterns has better noise robustness [35], we then displayed a set of Hadamard basis patterns on the DMD and projected them one-by-one onto the specimen through a tube lens (focal length: 200mm) and an objective lens. Subsequently, the spatial information of the specimen was encoded into time-varying light intensity signals. Finally, by using a 2D sensor (Pointgrey, GS3-U3-60QS6C-C, $1''$CCD, $2{,}736\times2{,}192$ pixels, pixel size: 4.54 μm), we simultaneously recorded the time-varying light intensity signals from different directions.

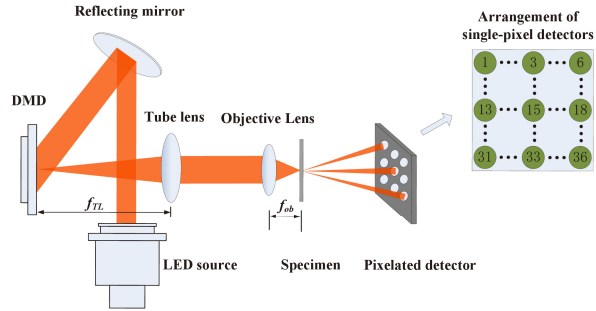

Fig. 3 Experimental setup of the proposed synthetic sampling microscope via single-pixel imaging.

### 4.1 Experimental result for the UASF1951 resolution target - sampling rate analysis

To verify the proposed method, we imaged a UASF1951 resolution target, where a $10\times$, NA 0.25 objective lens was used for illumination. The axial distance between the specimen and the 2D image sensor was approximately 8cm. According to these experimental parameters, the spatial cutoff frequency determined by the objective lens was $f_{ill}=0.25/0.633=0.395$ μm$^{-1}$.

Firstly, we carried out an experiment by only sampling the specimen from the illumination side. To realize it, we constructed a large size single-pixel detector by binning all the pixels of the camera together. If we project a set of Hadamard patterns (with an encoding pixel size of $86.4\mu m \times 86.4\mu m$), the spatial cutoff frequency determined by the Hadamard patterns was $f_{SLM}=10/(86.4\mu m)=0.116$ μm$^{-1}$. Obviously, $f_{SLM}<f_{ill}$; thus, this measurement system suffers from undersampling. Figure 4(a) shows an image ($64\times64$ pixels) reconstructed using this system, where we can only observe the feature of group 5, element 6 (resolution of 57.0 line pairs per mm). A straightforward method used to improve the resolution of the reconstructed image is the image-interpolation approach. With it, the low-sampling resolution image was fitted with a continuous function and then resampled with a finer interval. However, because no additional information is provided, the lost high-frequency details cannot be recovered using this method, as shown in Figs. 4(b) and 4(d). Therefore, the resolution of the retrieved image cannot be improved by using the image-interpolation method.

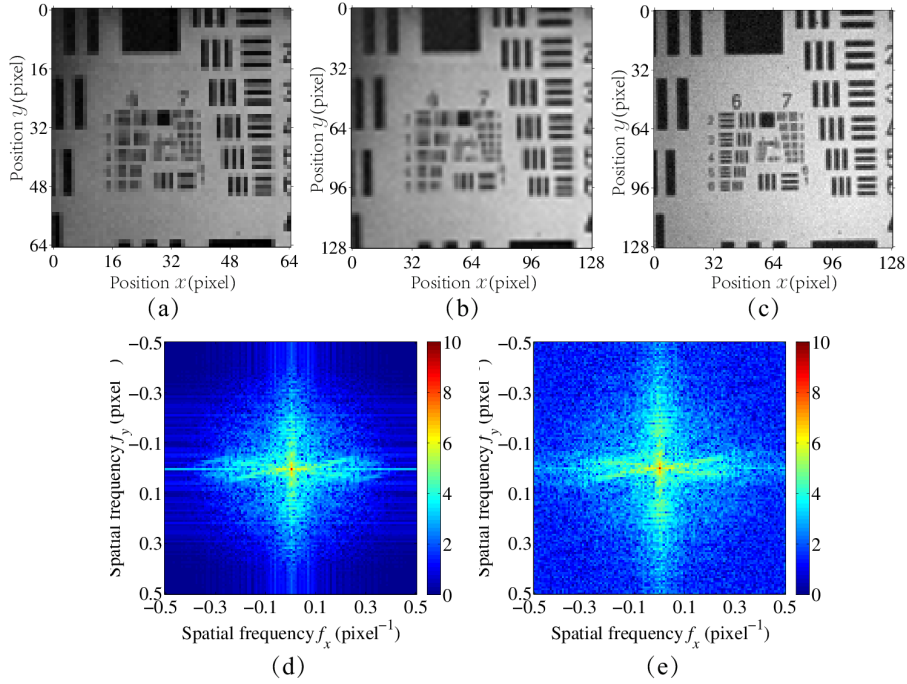

Fig.4. Images reconstructed by sampling the specimen only from the illumination side, all the pixels of the camera were binning together to form a large size single-pixel detector. (a) Image retrieved by projecting a set of Hadamard patterns (with an encoding pixel size of $86.4\mu m \times 86.4\mu m$); (b) image obtained by up-sampling Fig. 4(a) using the image-interpolation method; (c) image retrieved by projecting a set of Hadamard patterns (with an encoding pixel size of $43.2\mu m \times 43.2\mu m$); (d-e) Fourier spectrums corresponding to Figs. 4(b-c).

Next, to improve the resolution of the retrieved image, we increased the sampling rate of the illumination side. To achieve this, we employed a set of Hadamard patterns (with an encoding pixel size of $43.2\mu m \times 43.2\mu m$) to modulate the specimen. In the detection side, we still used only one large size single-pixel detector to sample the specimen. The large size single-pixel detector was constructed by binning all the pixels of the camera together. Figure 4(c) shows the reconstructed image, where the features of group 6, element 4 (resolution of 90.5 line pairs per mm) are resolved. Compared with Fig. 4(a), the sampling resolution of the reconstructed image shown in Fig. 4(c) was doubled. Additionally, the resolution improvement can be observed from their spectrum distributions (as shown in Figs. 4(d) and 4(e)). Note that for comparison, the Fourier spectrums shown in Figs. 4(d) and 4(e) were processed with the natural logarithm $\log(1+F)$, where $F$ is the Fourier spectrum. Unfortunately, Figure 4(c) was achieved at the cost of increasing the measurement time because the number of measurements for single-pixel imaging increases with the size of the reconstructed image. What's worse, reducing the pixel size of the Hadamard patterns will reduce the signal-to-noise ratio of the reconstructed image, as shown in Fig. 4(c). As a result, it might be difficult to obtain a high resolution and high quality image by simply increasing the sampling rate of the illumination sides.

To acquire a high resolution image with high signal-to-noise rate, we used the proposed SSI system to simultaneously sample the specimen from the illumination and detection sides. In the illumination side, we projected a set of Hadamard patterns (with an encoding pixel size of $86.4\mu m \times 86.4\mu m$). In the detection side, we used an array of $6\times 6$ single-pixel detectors to sample the specimen. Each single-pixel detector was constructed by binning $120\times 120$ adjacent pixels of the camera together. The lateral distance between the adjacent single-pixel detectors is 2.1 mm. To acquire multiple sub-pixel shifting images, we moved the specimen way from the focal plane of the objective lens with 40 μm. Then using the proposed system, we simultaneously retrieved 36 low-sampling rate images. Figures 5(a-d) show four images retrieved from the 1st, 6th, 31th, and 36th single-pixel detectors, respectively. The arrangement of the single-pixel detectors is shown in Fig. 3. Clearly, when the images were retrieved from different single-pixel detectors in the horizontal direction, the images of the specimen were horizontally shifted with respect to each other, as shown in Figs. 5(a) and 5(b). Similarly, when the images were retrieved from different single-pixel detectors in the vertical direction, the images of specimen were vertically shifted with respect to each other, as shown in Figs. 5(a) and 5(c). More importantly, as these low-sampling rate images were shifted with respect to each other via sub-pixel displacements, each image represented a different view of the same specimen, i.e., we can obtain new information in each low sampling rate image, as indicated by the red, blue, and green arrows in Fig. 5. The new information in each image can be exploited to synthesize a high-sampling rate image.

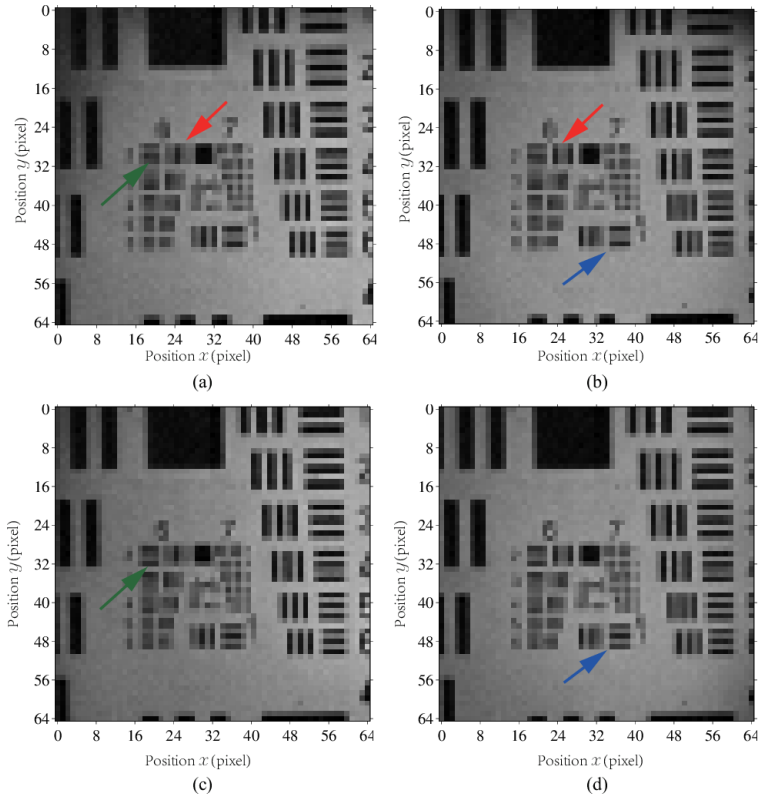

Fig.5. Low-sampling rate images retrieved using the proposed SSI system. (a-d) Images retrieved from the 1st, 6th, 31th and 36th single-pixel detectors, respectively. The arrangement of the single-pixel detectors is shown in Fig. 3.

To synthesize a high-sampling rate image from the low-sampling rate shifting images, we firstly calculated the sub-pixel shift of each low-sampling rate image by using the method presented in Section 3. The low-sampling rate image retrieved via the 1st single-pixel detector was chosen as the template image. Figure 6(a) shows a plot of the sub-pixel shifts between the template image and other low-sampling rate images. Using the calculated sub-pixel shifts, we successfully synthesized a high-sampling rate image shown in Fig. 6(b). Compared with the low-sampling rate image given in Fig. 5, the features of group 6, element 4 (resolution of

90.5 line pairs per mm) are now resolved in Fig. 6(b). Note, although the multiple single-pixel detectors are uniformly distributed on the $x-y$ plane, the sub-pixel shifts calculated and shown in Fig. 6(a) were not uniformly distributed on the $x-y$ plane. The irregular distribution of the sub-pixel shifts might be caused by many factors, finding out the reason is the further work we will focus on.

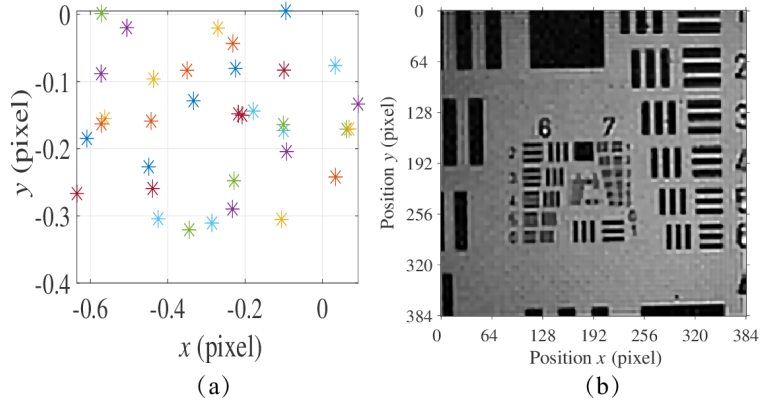

Fig.6. (a) Sub-pixel shifts calculated by the method mentioned in Section 3; (b) high-sampling rate image synthesized by using the low-sampling rate images and the sub-pixel shifts shown in Fig. 6(a).

Note we used an array of 6×6 single-pixel detectors to shift the low sampling rate images by one sixth of a pixel in the horizontal and vertical directions, respectively. Accordingly, compared with the low sampling rate images, the horizontal and vertical magnification factors of the high sampling rate image are set as 6. In other words, the resolution of the high sampling rate image is 384×384 pixels. From Fig. 6(b), we can see the features of group 6, element 4 (resolution of 90.5 line pairs per mm) are now resolved. Clearly, the resolving capability shown in Fig. 6(b) is the same as that shown in Fig. 4(c). This indicates that the sampling rate of the imaging system can be doubled by using the proposed SSI method. More importantly, because all the single-pixel detectors could operate simultaneously, compared with existing sampling rate enhancement methods [26-29], the sampling resolution of the reconstructed image was improved without mechanical scanning or increasing the measurement time.

### 4.2 Imaging of biological specimen

To demonstrate the performance of the proposed method for a biological specimen, we imaged a slide of a banyan pillar with a 10×, NA 0.25 objective lens. When we only used one single-pixel detector in the detection side, the sampling rate of the imaging system was determined by the illumination side. For example, when we projected a set of Hadamard patterns (with an encoding pixel size of 43.2 μm×43.2 μm) onto the specimen, we could not observe the small structure of the specimen highlighted by the dashed line in the reconstructed image (128×128 pixels), as shown in Fig. 7(a). When we increased the sampling rate of the illumination side by projecting a set of Hadamard patterns (with an encoding pixel size of 21.6 μm×21.6 μm), the sampling resolution of the reconstructed image shown in Fig. 7(b) was doubled. However, the signal-to-noise ratio of the reconstructed image was reduced because the light available for each pixel of Hadamard patterns was reduced by half. Even worse, the measurement time for Fig. 7(b) was 4 times longer than that for Fig. 7(a) because the image size of Fig. 7(b) is 256×256 pixels.

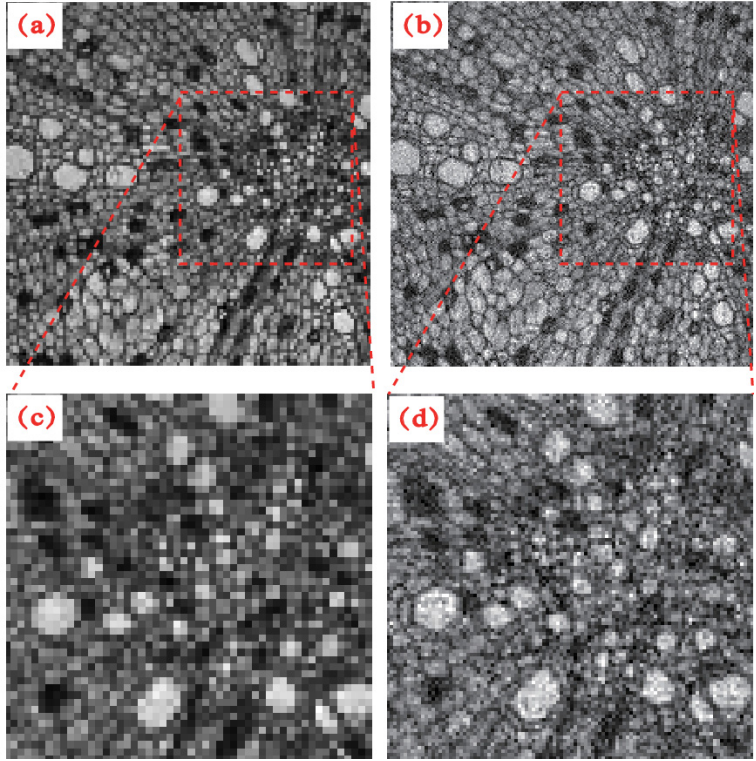

Fig. 7. Images of the biological specimen retrieved by sampling it from the illumination side. (a) Low-sampling rate images (128×128 pixels) retrieved by projecting a set of Hadamard patterns (with an encoding pixel of 43.2 μm×43.2 μm ); (b) image ( 256×256 pixels) retrieved by projecting a set of Hadamard patterns (with an encoding pixel of 21.6 μm×21.6 μm ); (c-d) partial enlarged views of Figs. 7(a-b)

To overcome the aforementioned drawbacks and acquire a high-sampling rate image, we used the proposed SSI system to simultaneously sample the specimen from the illumination and the detection sides. To sample the specimen from the illumination side, we projected a set of Hadamard patterns (with an encoding pixel size of 43.2 μm×43.2 μm ) onto the specimen. For the detection side, we used an array of 6×6 single-pixel detectors to sample the specimen. Besides, to obtain multiple shifting images, the specimen was placed away from the focal plane of the objective lens with 20 μm . Then, we simultaneously retrieved 36 low-sampling rate images using the proposed SSI system. Figures 8(a-d) show four images retrieved from the 1st, 6th, 31th, and 36th single-pixel detectors, respectively. In contrast to Fig. 5, because the axial distance between the specimen and the focal plane of objective was smaller, the shift amounts between Figs. 8(a-d) are too small to be observed by the naked eye. However, using the method presented in Section 3, the sub-pixel shifts for all the images can be acquired, as shown in Fig. 9(a). According to the sub-pixel shifts and all the low-sampling rate images, we synthesized a high-sampling rate image shown in Fig. 9(b), where its partial enlarged view is presented in Fig. 9(c). Compared with the image shown in Fig. 7(a) retrieved by sampling specimen only from the illumination side, the details highlighted by the dashed line are resolved in Fig. 9(c). These experimental results demonstrate that the proposed SSI method is effective for biological imaging.

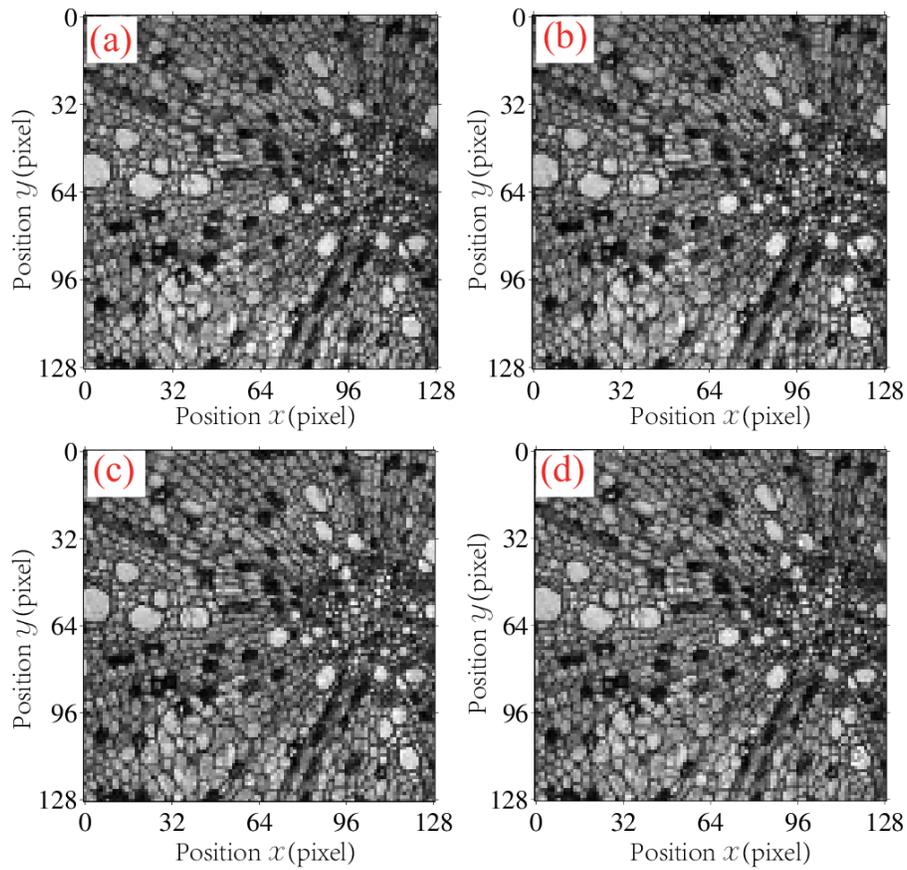

Fig.8. Low-sampling rate images of the biological specimen retrieved by using the proposed SSI system. (a-d) Images retrieved from the 1st, 6th, 31th, and 36th single-pixel detectors, respectively.

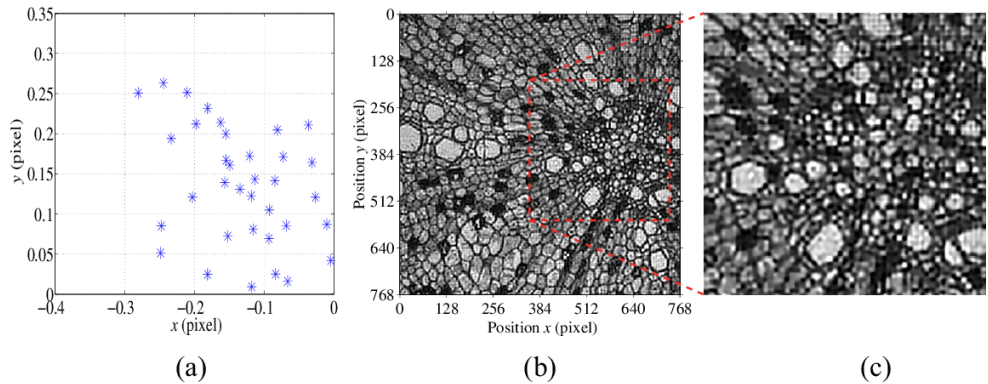

Fig. 9. Results of biological specimen. (a) Sub-pixel shifts of all the low-sampling rate images. (b) high-sampling rate image reconstructed by simultaneously sampling the specimen from the illumination and the detection sides; (c) partial enlarged view of Figs. 9(b).

### 4.3 Effect of the detection sampling rate on reconstruction quality

In this section, we will show how the sampling rate of the detection side affected the reconstruction quality. This investigation was performed by imaging a resolution target (as used in Subsection 4.1) under different detection sampling rates and synthesizing a high sampling rate image using the proposed SSI system. The changes of the sampling rate of the detection side was realized by using arrays of 2×2, 4×4, 6×6 and 8×8 single-pixel detectors in experiment, where the lateral distances between the adjacent single-pixel detectors were set as 5.9 mm, 3.1 mm, 2.1 mm, and 1.5 mm, respectively. The defocusing distance of the specimen was 40 μm. The other experimental parameters of the imaging system were the same as those used in Subsection 4.1.

The experimental results obtained with different detection sampling rates are shown in Fig. 11, where Figs. 10(a1-a4) present the sub-pixel shift plots of all the low-sampling rate images by using different detection sampling rates; Figs. 10(b1-b4) present four high-sampling rate images by using different detection sampling rates, and Figs. 10(c1-c4) show the Fourier spectrum corresponding to Figs. 10(b1-b4). Note that for comparison, the sizes of all the four high-sampling rate images were set as 384×384 pixels, and the four Fourier spectrums were processed with the natural logarithm $\log(1+F)$, where $F$ is the Fourier spectrum. Besides, in each group of the low-sampling rate images, the image retrieved from the upper-left single-pixel detector was set as template image. The sub-pixel shift plots given in Fig. 10(a1-a4) are the shifts between the template image and other low-sampling rate images.

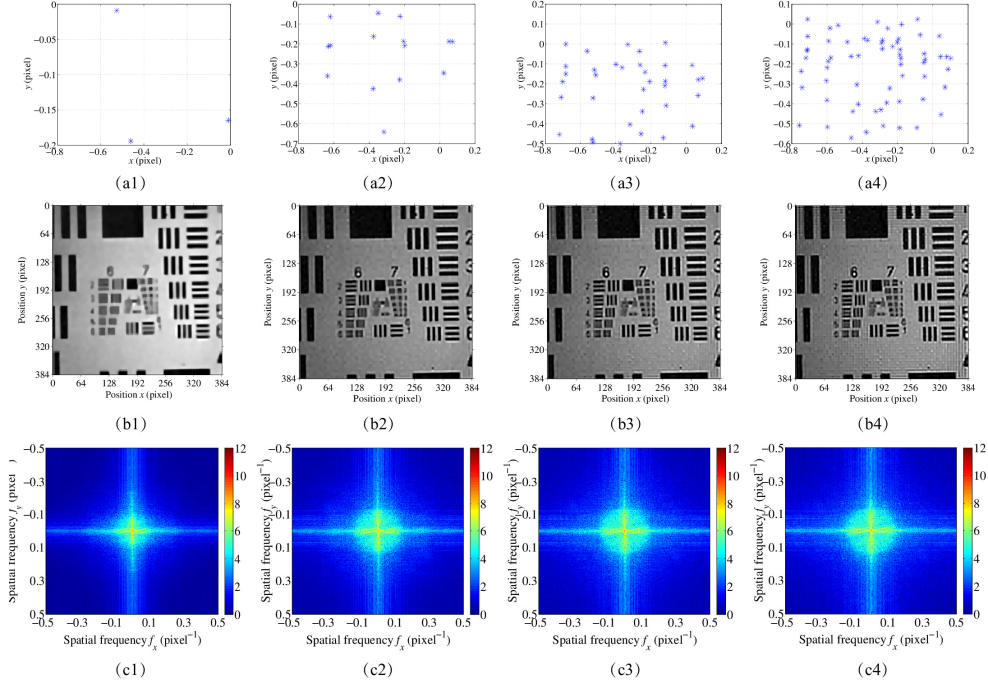

Fig. 10. Results obtained by setting the sampling rate of the detection side as 2×2, 4×4, 6×6, and 8×8, respectively. (a1-a4) Sub-pixel shifting amounts of the low-sampling rate images ; (b1-b4) high-sampling rate images; (c1-c4) Fourier spectrums corresponding to Figs. 10(b1-b4).

As shown in Figs. 10(a1), 10(b1) and 10(c1), when the sampling rate of the detection side was 2×2, we can only observe the features of group 5, element 6 (resolution of 57.0 line pairs per mm) in the high-sampling rate image. Compared with the low-sampling rate image shown in Fig. 5, the resolving capability of the high-sampling rate image was almost not improved. When we increased the sampling rate of the detection side to 4×4, we can clearly observe the features of group 6, elements 3 (resolution of 80.6 line pairs per mm) in the high-sampling rate image shown in Fig. 10(b2). The Fourier spectrum shown in Fig. 10(c2) also indicates that the image of Fig. 10(b2) has higher resolution than Fig. 10(b1). When the sampling rate of the detection side was increased to 6×6, we can clearly observe the feature of group 6, element 4 (resolution of 90.5 line pairs per mm) in the high-sampling rate image shown in Fig. 10(b3). The Fourier spectrum shown in Fig. 10(c3) also indicates that resolution of Fig. 10(b3) is slightly higher than Fig. 10(b2). But when the sampling rate of the detection side was increased to 8×8, we still only observe the feature of group 6, element 4 (resolution of 90.5 line pairs per mm) in the high-sampling rate image shown in Fig. 10(b4). The Fourier spectrum shown in Fig. 10(c4) also indicates the resolution of Fig. 10(b4) is almost the same as that of Fig. 10(b3). These results indicate that the synthesis of the illumination sampling and detection sampling will improve the resolution of the reconstructed image, but when the sampling rate of the detection side exceeds 6×6, the resolution improvement becomes marginal. This conclusion agrees with the limit analysis of the sub-pixel shifting algorithm given in reference [36].

## 5. Discussion

Here to retrieve multiple sub-pixel shifting images, the specimen should be moved away from the focal plane of the objective lens with a small distance. But it should also be emphasized that this will not reduce the resolution of the sub-pixel shifting images. Because the Hadamard patterns appear in focus within the depth of field of the objective lens as shown in Fig. 11. As such, if the specimen is moved away from the focal plane but within the depth of field of the objective lens, the specimen can still be accurately modulated by the Hadamard pattern and therefore the resolution of the retrieved image will not be reduced.

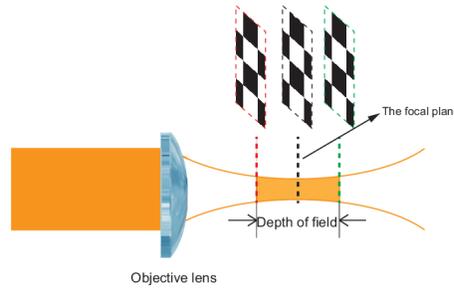

Fig. 11. Depth of field of objective lens.

To validate the analysis given above, we carried out an experiment by slightly moving the specimen away from the best focus of the objective lens with different distances. We used a 10×, NA 0.25 objective lens to project the Hadamard patterns (the encoding pixel size of the Hadamard patterns is 86.4 μm×86.4 μm). The wavelength of the illumination source is 0.633 nm. Based on these parameters and using the method mentioned in [37], the depth of field of the objective lens can be estimated as 44.688 μm ($\frac{0.633}{0.25^2}+\frac{1\times86.4}{10\times0.25}=44.688$ μm).

The test specimen was a USAF resolution target. Each single-pixel detector was constructed by binning 120×120 adjacent pixels of the camera together. The other experimental parameters were the same as that used in Subsection 4.2. At first, the specimen was placed at the focal plane of the objective lens. Figure 12(a) shows the image retrieved at the focal plane. Next, the specimen was slightly moved away from the focal plane of the objective lens with 20 μm and 40 μm, respectively. The corresponding images retrieved at the two positions are shown in Figs. 12(b-c). Compared with the image retrieved at the focal plane of the objective lens, the resolution of the images shown in Figs. 12(b-c) are almost not reduced. Note that to retrieve sub-pixel shifting images for synthesizing the high-resolution image shown in Fig. 6(b), we only need to move the specimen from the focal plane of the objective lens with

40 μm. In this case, the specimen is still located within the depth of field of the objective lens and therefore the resolution of the sub-pixel shifting image will not be reduced. From the these experimental results, we can conclude that slightly moving the specimen from the focal plane but within the depth of field of the objective lens will not reduce the resolution of the retrieved image. Instead, it allows us to retrieve multiple sub-pixel shifting images for the synthesis of high-resolution image.

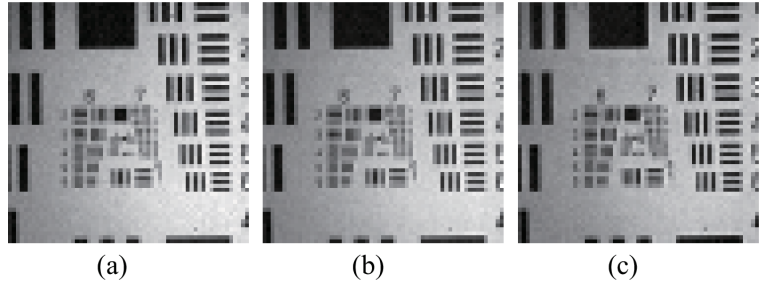

(a)            (b)            (c)

Fig. 12. (a) Image retrieved by placing the specimen at the focal plane; (b-c) Images retrieved by moving the specimen away from the focal plane with 20 μm and 40 μm, respectively

## 6. Conclusion

We propose a concept of synthetic sampling imaging and demonstrate it experimentally via an example of microscopy. The proposed system uses a 2D sensor and an SLM so as to simultaneously sample specimen from the illumination and detection sides. Based on the principle of single-pixel imaging, we can retrieve multiple sub-pixel shifting images from multiple photosensitive units of the 2D sensor, without requiring mechanical scanning. Using these sub-pixel shifting images, we can double the sampling resolution of the reconstructed image and therefore alleviate the undersampling problem of 2D sensor or SLM. The proposed sampling scheme, where the sampling task is shared by the illumination and detection sides, provides an insight to develop a high sampling rate imaging system. This is particularly important for non-visible light imaging system because fabricating high resolution 2D sensor or SLM that can operate in non-visible light spectrum remains difficult or very expensive.

## Appendix

### A. Steps used to determine the weighting matrix P [30]

According to Eq. (8), each pixel of the low-sampling rate image equals to the inner product of a row vector of the weighting matrix $P$ and all the pixels of the high-sampling rate image, where each weighting coefficient corresponds to a high-sampling rate pixel. On the other side, each low-sampling rate pixel is mainly determined by its near neighbor high-sampling rate pixels. The closer distance between the low-sampling and the high-sampling rate pixels, the larger the weighting coefficient. As such, the weighting coefficients can be approximately by a Gaussian distribution [30]. Assume that we obtained $n$ low-sampling rate images, each with size of $(N_1, N_2)$. Then the steps used to build the weighting matrix $P$ can be summarized as shown in Fig. 13. (1) Determine the horizontal and vertical magnification factors $(L_1, L_2)$ between the high-sampling rate image and the low-sampling rate images. For example, if the number of the horizontal shifts is 6, $L_1$ should not be larger than 6. Based on the magnification factors, we can generate a high-sampling rate grid $G_h$ with size of $(L_1 N_1, L_2 N_2)$. (2) Transform all the pixels of each low-sampling rate image into $G_h$ by using the calculated sub-pixel shifts and magnification factors. Clearly, the new pixel coordinates of the low-sampling rate images do not coincide with the integer grid but scatter in $G_h$. (3) Calculate the pixel distance (denote as $d_{k,i,j}$) between each pixel of the low-sampling rate image and each integer grid point of $G_h$, where $k\ (k=1,2,\cdots,n)$ represents the order of the low-sampling rate image, $i\ (i=1,2,\cdots,N_1N_2)$ represents the order of the point in the $k$th low-sampling rate image, and $j\ (j=1,2,\cdots,L_1N_1L_2N_2)$ represents the order of the grid point in $G_h$. (4) Determine the standard deviation $\sigma\ (\sigma\in[0,1])$ of the Gaussian distribution function. (5) Calculate the coefficients $P_{k,i,j}$ of the weighting matrix one-by-one using the following expression,

$$P_{k,i,j} = e^{\left(-\frac{d_{k,i,j}^2}{2L_1\times L_2\times \sigma^2}\right)}. \tag{9}$$

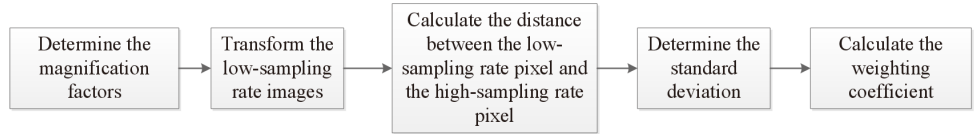

Fig. 13. Flowchart used to acquire the weighting matrix *P*.

## Funding